\documentclass[11pt]{amsart} 
\usepackage{graphicx}
\usepackage{amssymb, amsmath}
\usepackage{amsfonts}
\usepackage{amssymb}
\usepackage{float}

\newcommand{\R}{\mathbb R}

\def\be{\begin{equation}}
\def\ee{\end{equation}}

\def\R{\mathbb R}

\begin{document}
\bibliographystyle{siam}

\title[Rayleigh-B\'enard Convection]{Addendum to the paper  "Two-Dimensional Infinite Prandtl Number Convection: 
Structure of Bifurcated Solutions, J. Nonlinear Sci., 17(3), 199-220, 2007"}
\author[Ma]{Tian Ma}
\address[TM]{Department of Mathematics, Sichuan University,
Chengdu, P. R. China and Department of Mathematics,
Indiana University, Bloomington, IN 47405}

\author[Park]{Jungho Park}
\address[SW]{Department of Mathematics,
Indiana University, Bloomington, IN 47405}
\email{junjupar@indiana.edu}

\author[Wang]{Shouhong Wang}
\address[SW]{Department of Mathematics,
Indiana University, Bloomington, IN 47405}
\email{showang@indiana.edu}
\thanks{The work was supported in part by the
Office of Naval Research  and by the National Science Foundation.}


\maketitle


The main objective of this addendum to the mentioned article \cite{park2} by Park is  to provide some  remarks on bifurcation theories for nonlinear partial differential equations (PDE) and their applications to fluid dynamics problems. We only wish to comment and list some related literatures, without any intention to  provide a complete survey. 

For steady state PDE bifurcation problems, the often used classical bifurcation methods include  1) the Lyapunov-Schmidt procedure, which   reduces the PDE problem to a finite dimensional algebraic system, 2) the Krasnoselskii theorem for bifurcations crossing an eigenvalue of odd algebraic multiciplicity \cite{krasnoselski} , 3) the Krasnoselskii theorem for potential operators, 4) the Rabinowitz global bifurcation 
theorem  \cite{rabinowitz71},  5) Crandall and Rabinowitz  theorem for bifurcations crossing a simple eigenvalue \cite{CR71}, and 6) bifurcation from higher-order terms, regardless of the multiplicity of the eigenvalues \cite{mw04e,mw04f}. We also refer the interested readers to, among many others,  \cite{nirenberg,ch, GS1,GS2, b-book, chinese-book} for more comprehensive discussions.
Nirenberg have a beautiful survey paper \cite{nirenberg81} on topological and variational methods for nonlinear problems, which has influenced a whole generation of nonlinear analysts. 

The Hopf bifurcation, also  called Poincar\'e-Andronov-Hopf bifurcation, was  independently studied and discovered by Andronov in 1929 and Hopf in 1942 and Poincar\'e  in 1892  for ordinary differential equations.    In particular, in  his paper  
\cite{Hopf42}, Hopf also indicated the possible application of the Hopf bifurcation theorem  to  bifurcation of time  periodic solutions for the Navier-Stokes equations. The Hopf bifurcation  was generalized to infinite dimensional setting for PDEs by Crandall and  Rabinowitz \cite{CR77}, Marsden and McCracken \cite{MM76}, and Henry \cite{henry}. We mention in particular the last two references using the center manifold reduction procedure to reduce the problem to a finite dimensional problem. 


Over the last 30 years or so, there have been extensive studies using the bifurcation theory with symmetry methods and applications; see, among many others,  
D. Sattinger \cite{Sattinger78, Sattinger79, Sattinger80, Sattinger83}, 
M. Golubitsky, I. Stewart, and D. Schaeffer \cite{GS1, GS2}, and M. Field \cite{field}.



Recently, Ma and Wang have developed a bifurcation theory for nonlinear 
PDEs \cite{mw-db1, mw-benard}. This bifurcation theory 
is centered at a new notion of bifurcation, called attractor bifurcation for 
nonlinear evolution equations, and is synthesized in two recent books 
by Ma and Wang \cite{b-book, chinese-book}. Furthermore, this new bifurcation theory has been further developed by Ma and Wang into a complete new dynamic transition theory for nonlinear problems; see two recent books by Ma and Wang 
\cite{chinese-book, ptd} and the references therein for a more detailed account of the theory. These new theories
has been used in many problems in sciences and engineering, including the B\'enard convection problem and the Taylor problem in the classical fluid dynamics, 
doubly-diffusive convections and rotating Boussinesq equations in geophysical fluid dynamics, many phase transition problems in statistical physics, 
biology and chemistry; see 
 \cite{mw05g,mw05a,mw05c, b-book, chinese-book, ptd}  and the references therein.

\begin{figure}
        \centering \includegraphics[width=0.6\hsize]{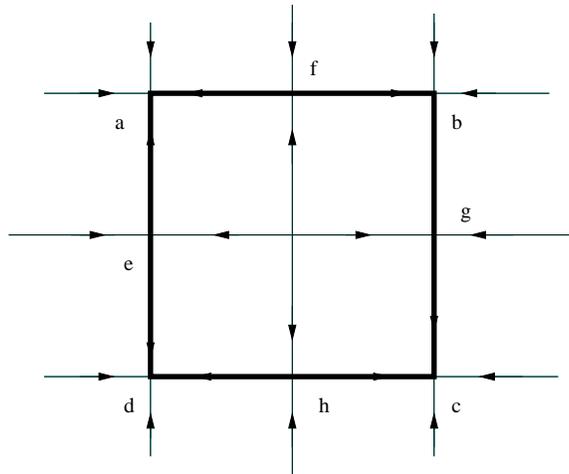}
        \caption{A bifurcated attractor containing 4 nodes (the points a, b, c, and d), 4 saddles (the points e, f, g, h), and orbits connecting these $8$ points.}
\label{fg4.1}
\end{figure}
The main purpose of this new bifurcation theory is to study  transitions of one state to another in nonlinear problems. 
We illustrate the concept by a simple example. 
For $x =(x_1, x_2) \in \R^2$, the system $\dot x = \lambda x - (x_1^3, 
x_2^3)$
bifurcates from $(x, \lambda)=(0,0)$ to an attractor  $\Sigma_\lambda=S^1$. 
This bifurcated attractor is as shown in Figure \ref{fg4.1}, and contains 
exactly 4 nodes (the points a, b, c, and d), 4 saddles (the points e, f, 
g, h), and orbits connecting these $8$ points.
{\it From the physical transition point of view, as $\lambda$ crosses $0$, the new state after the system undergoes a transition is represented by the {\bf whole bifurcated attractor} $\Sigma_\lambda$, rather than any of the steady states or any of the connecting orbits.} 
Note that the global attractor  is the 2D region enclosed by $\Sigma_\lambda$.
We point out here that the bifurcated attractor is different from the study on global attractors of a dissipative dynamical system-both finite and infinite dimensional. Global attractor studies the global long time dynamics (
see among others \cite{FT79, BV83, CF85, CFT85}), while the bifurcated attractor provides a natural object for studying dynamical transitions \cite{b-book, chinese-book, ptd}.

\bigskip

Now we return to the Rayleigh-B\'enard convection problem. This is classical problem in fluid dynamics. The study for this problem on the one hand plays an important role in 
understanding the turbulent behavior of fluid flows, 
and on the other hand often leads to new insights and methods toward 
solutions of other problems in sciences and engineering. 

Linear theory of the Rayleigh-B\'enard problem were essentially derived by physicists; see, among others, Chandrasekhar \cite{chandrasekhar} and  Drazin and Reid \cite{dr}.
Bifurcating solutions of the nonlinear problem were first constructed formally 
by Malkus and Veronis \cite{veronis58}. The first rigorous proofs of the existence of 
bifurcating solutions were given by Yudovich \cite{yudovich67a, yudovich67b} and 
Rabinowitz \cite{rabinowitz}. Yudovich  
proved the existence of bifurcating solutions by a topological degree argument. Earlier, however, Velte \cite{velte} had proved the existence of branching 
solutions of the Taylor problem by a topological degree argument as well. 
The application of group-theoretic bifurcation theory to the B\'enard convection 
are explored in last thirties years or so by many authors.   To the best knowledge of authors, Sattinger's papers \cite{Sattinger77, Sattinger78, Sattinger80} are the first ones to use the group-theoretic point of view to study this problem in combination with the Lyapunov-Schmidt reduction procedure. Then group-theoretic methods are  used,  in conjunction with center manifold reduction and leading to the amplitude equations, to fluid problems by, among others,   M. Golubitsky, I. Stewart, and D. Schaeffer 
\cite{GS2},  Chossat and  his collaborators \cite{CI94,CL00}, Iooss and his collaborators \cite{IA98,VI92}, and the references therein. We would like also to mention the Hopf bifurcation result obtained by Chen and Price \cite{chen-z-m}, where they use the continued fractions method first introduced by Meshalkin  and Sinai \cite{sinai}.
Recently, Ma and Wang have used their newly developed attractor bifurcation theory mentioned earlier to study the B\'enard convection problem. In \cite{mw-benard}, they proved that the Rayleigh-B\'enard  problem bifurcates from the basic state to an 
attractor $A_R$ when the
Rayleigh number $R$ crosses the first critical Rayleigh number $R_c$ 
for all physically  sound boundary conditions, regardless of {\it the geometry of the domain and the  multiplicity of the eigenvalue $R_c$ for the linear problem}. 
Furthermore detailed characterization of solutions in the bifurcated attractor $A_R$ in both the physical  and the phase spaces for special geometries of the domain is given in \cite{mw07}. Also,  the bifurcated attractor  $A_R$ attracts {\bf any bounded set} in 
$H\setminus \Gamma$, where $H$  is the whole phase space and $\Gamma$ is the stable manifold of the basic solution.  
Finally, we would like to point out  that the bifurcation and stability analysis discussed in this and  the above mentioned articles are for viscous flows, and we refer to 
Friedlander  and Yudovich  \cite{FY99} and 
Friedlander,  Strauss,  and Vishik \cite{FSV97} and the references therein for  details.

\medskip

The work presented in Park \cite{park1, park2} is an application of the aforementioned new attractor bifurcation theory to the infinite Prandtl number B\'enard convection. 
Here are some specific comments on the results in these articles and some further developments.

\medskip

{\sc First}, the results obtained are motivated by the  attractor bifurcation theory 
\cite{b-book} and the geometric theory for incompressible flows \cite{amsbook}, both developed recently by Ma and Wang \cite{b-book}. Without the new insights from these theories, one does not come up with the theorems proved in \cite{park1, park2} and in other related articles, as evidenced by the fact that no such theorems have been stated in the vast existing literature on the Rayleigh-B\'enard convection.

\medskip

{\sc Second}, the types of solutions in this $S^1$ attractor depend on the boundary conditions. With the periodic boundary condition  in the $x_1$ direction in this article, the bifurcated attractor consists of only steady states. 

In fact, when  the boundary conditions for the velocity field  are free slip boundary conditions  and the spatial domain is $\Omega=(0, L)^2\times (0, 1)$ with $0 < L^2 < (2-\sqrt[3]{2})/(\sqrt[3]{2}-1)$,  Ma and Wang \cite{mw07, ptd}  prove that the bifurcated attractor is still an $S^1$, consisting of exactly eight singular steady states (with four saddles and four minimal attractors) and eight heteroclinic orbits connecting these steady states. The bifurcated attractor and its detailed classification provide a  global dynamic transition in both the physical and phase spaces.

Furthermore, again in a more general three-dimensional (3D) domain 
$\Omega=(0, L_1)\times (0, L_2)\times (0, 1)$, 
with doubly-periodic boundary conditions in the horizontal directions and the free-boundary conditions on the top and bottom, it is proved  \cite{ptd} that the bifurcated attractor is 
either $\Sigma_R =S^5$ if $L_2=\sqrt{k^2-1} L_1$  for $k=2, 3, \cdots$, or else
$\Sigma_R =S^3$. {\it It is clear that these bifurcated attractors contains many more solutions than the  solutions  derived by any group-theoretic methods.}
 
Hence, we iterate here that  the method and ideas developed by Ma and Wang are crucial to obtain these results, which can not be obtained using only the classical bifurcation theories. For the case studied in this article, the classical bifurcation theory with symmetry arguments implies that  the bifurcated attractor  {\it contains} a circle of steady states. We need, however,  the new bifurcated theory to prove in particular that the bifurcated attractors are {\it exactly} an $S^1$. 

Furthermore, it is also much obvious, as partially indicated,  that the 3D results can not be derived by group-theoretic methods.  In addition, for general boundary conditions such as the free-slip boundary conditions mentioned above, no symmetry can be used, and the classical amplitude equation methods fails to derive the dynamics.

In short, although group-theoretic bifurcation methods are useful in different problems, the need for other more general methods is obvious and inevitable as suggested by, among others,
D. Sattinger \cite{Sattinger78,Sattinger80}, one of the pioneers of the group-theoretic bifurcation methods, and  B. Gr\"unbaun \cite{grunbaun}.

\medskip

{\sc Third}, it is proved in \cite{mw-benard, mw07, park1, park2} that the bifurcated attractor in the B\'enard convection problem attracts {\bf any bounded set} in 
$H\setminus \Gamma$, where $H$  is the whole phase space and $\Gamma$ is the stable manifold of the basic solution.  To the best of the knowledge of authors, this "global" stability result can not be derived from any existing methods. It is obtained by using a new stability result proved by Ma and Wang \cite{mw-benard}, which is derived using a combination of energy estimates and topological arguments.

As Kirchg\"assner indicated in  \cite{kirch}, 
"an ideal stability theorem would include all physically meaningful 
perturbations and establish the local stability of a selected class of 
stationary solutions, and today we are still far from this goal."
On the other hand, fluid flows are normally time dependent. Therefore  
bifurcation analysis for steady state problems provides in general 
only partial answers to the problem, and is not enough for solving 
the stability problem. Hence it appears that the right notion of 
asymptotic stability after the first bifurcation should be best described 
by the attractor near, but excluding, the trivial state. It is one of 
our main motivations for introducing attractor bifurcation theory and the dynamic transition theory \cite{b-book, chinese-book, ptd}.

\medskip

{\sc Fourth}, the geometric theory for incompressible flows recently developed by 
Ma and Wang \cite{amsbook} is crucial for  the structure and its stability of the solutions in the physical spaces obtained in the main theorems in \cite{park2}. 
Also, we note that  a special  structure with rolls separated by a cross channel flow derived 
in \cite{mw07} has not been rigorously examined in the B\'enard convection setting although it has been  observed in other physical contexts such as the Branstator-Kushnir waves in the atmospheric dynamics 
\cite{branstator,kushnir}.

For completeness, we mention that this geometric theory for incompressible flows   consists of research in two steps: 1) the study
of the structure and its transitions/evolutions of divergence-free
vector fields, and 2) the study of the structure and its
transitions of velocity fields for 2-D incompressible fluid flows
governed by the Navier-Stokes equations or the Euler equations.
The original motivation of this research program was 
to understand the dynamics of the ocean currents 
in the physical space, and  it turns out that there is 
a much richer new mathematical theory with more applications 
than the original motivation from the Oceanography. 
Among other results, for example, 
the theory leads to a  rigorous characterization 
of boundary layer separation on when, where, and how the 
separation occurs and to make connections between 
the time and location of the separation; see Ghil, Ma and Wang \cite{gmw1, gmw2}, 
Ma and Wang \cite{amsbook}, and the references therein.
This is a long standing problem in fluid 
mechanics going back to the pioneering work of Prandtl (1904); 
see also Chorin and Marsden \cite{chorin}, and  J{\"a}ger,  Lax and Morawetz \cite{jlm}.

With this characterization of the boundary layer-separation in our disposal, 
Ma and Wang \cite{mw07b} are able to derive a rigorous characterization of the boundary-layer and interior separations in the Taylor-Couette-Poiseuille flow. The results obtained  provide a rigorous characterization on how, when and where the propagating Taylor vortices (PTV) are generated. 
In particular, contrary to what is commonly believed, it is shown that the 
PTV do not appear after the first dynamical bifurcation, and they appear only when the Taylor number is further increased to cross another critical value so that a structural bifurcation occurs. This structural bifurcation corresponds to the boundary-layer and interior  separations of the flow structure in the physical space.  
\bibliography{addendum}
\end{document}